# Low-Resource Spoken Language Identification Using Self-Attentive Pooling and Deep 1D Time-Channel Separable Convolutions

**Roman Bedyakin**
AO HTSTS, Moscow, Russia
rbedyakin@ntr.ai

**Nikolay Mikhaylovskiy**
NTR Labs, Moscow, Russia
Higher IT School of Tomsk State
University, Tomsk, Russia
nickm@ntr.ai

## Abstract

This memo describes NTR/TSU winning submission for Low Resource ASR challenge at Dialog2021 conference, language identification track.

Spoken Language Identification (LID) is an important step in a multilingual Automated Speech Recognition (ASR) system pipeline. Traditionally, the ASR task requires large volumes of labeled data that are unattainable for most of the world's languages, including most of the languages of Russia. In this memo, we show that a convolutional neural network with a Self-Attentive Pooling layer shows promising results in low-resource setting for the language identification task and set up a SOTA for the Low Resource ASR challenge dataset.

Additionally, we compare the structure of confusion matrices for this and significantly more diverse VoxForge dataset and state and substantiate the hypothesis that whenever the dataset is diverse enough so that the other classification factors, like gender, age etc. are well-averaged, the confusion matrix for LID system bears the language similarity measure.

**Keywords:** low-resource languages, spoken language identification, convolutional neural networks, self-attentive pooling
**DOI:** 10.28995/2075-7182-2021-20-XX-XX

## 1 Introduction

Spoken Language Identification (LID) is a process of classifying the language spoken in a speech recording [43] and is an important step in a multilingual Automated Speech Recognition (ASR) system pipeline.

Differences between languages exist at all linguistic levels and vary from marked, easily identifiable distinctions (such as the use of entirely different words) to more subtle variations (e.g. the use of aspirated vs. unaspirated syllable-initial plosives in English vs. French). The latter end of the range is a challenge not only for automatic LID systems but also for linguistic sciences themselves [23]. The fact that differences between varieties of a single language may be larger than those between two distinct languages are important when developing LID for the low-resource languages existing in a context of prevalent major language (for example, Russian).

In this memo, we show that a convolutional neural network with a Self-Attentive Pooling layer shows promising results in low-resource setting for the language identification task. Our system detects the language, the genus and the family for an utterance, although, unlike the original competition task requirement, we do not discern the surprise language utterances, so our system tries to classify the utterance into the languages present in the training set only. The system described herein is similar to the one simultaneously submitted for SIGTYP 2021 Shared Task on predicting language IDs from speech, although the dataset is completely different.

### 1.1 Previous work

The first works on LID date back at least to mid-seventies, when Leonard and Doddington explored frequency of occurrences of certain reference sound units in different languages [17] [18].

Previously developed LID approaches include:

- Purely acoustic LID that aims at capturing the essential differences between languages by modeling distributions in a compact representation of the raw speech signal directly [18] [40].
- Phonotactics LID rely on the relative frequencies of sound units (phoneme/phone) and their sequences in speech [18] [40].
- Prosodic LID use tone, intonation and prominence, typically represented as pitch contour [18] [40].
- Word Level LID systems use fully-fledged large vocabulary continuous speech recognizers (LVCSR) to decode an incoming utterance into strings of words and then use Written Language Identification.

In the latest 10 years, intermediary-dimensional vector representations similar to i-vector [7][8][13] and x-vector [35] have been dominating the speech classification field, including LID. Additionally, starting from, likely, 2014 [19], deep neural networks have been predominantly used for such tasks (see, for example, [1][34][20]), although the first applications of deep neural networks to LID date back to 2009 [22].

For years, LID technology evaluation have been driven by US National Institute of Standards and Technology (NIST). The LRE17 [33] was the latest in the ongoing series of language recognition technology evaluations conducted by NIST since 1996 [24]. Since LRE11 [25], the focus of the language detection task was in differentiating closely related languages that are sometimes mutually intelligible, as the overall technology was considered mature enough to distinguish distant languages robustly.

In a somewhat parallel track, LID technology evaluations were performed at Albayzin Language Recognition Evaluations held in 2008 [29], 2010 [30] and 2012 [31] and organized by Software Technology Working Group of the University of the Basque Country, with the support of the Spanish Thematic Network on Speech Technology. Unfortunately, neither NIST LRE datasets no KALAKA-3 dataset of Albayzin LRE are not publicly/freely available, so it is hard to benchmark our approach against the ones evaluated in these competitions.

In 2021, we see a new surge of research interest to LID, especially in the context of low-resource languages [20][36][20][36]. One of the reasons is that for many low-resource and endangered languages, only single-speaker recordings may be available, indicating a need for domain and speaker-invariant language ID systems [36]. Thus, generalization to different speakers, domains and conditions becomes important in the LID system design.

For example, Abdullah et. al in [1] address the following research questions:
- RQ1: To what degree do neural LID models for related languages generalize to another domain with different acoustic conditions?
- RQ2: Are different low-level speech features equally robust under domain mismatch?
- RQ3: Can we adapt LID models to a new domain without using labelled data in the new domain? If yes, what are the factors that affect the adaptability of the model?

They find that generalization from the radio broadcast speech to read speech is better than in the reverse direction.

van der Merwe [20] proposes a mix of triplet and cross-entropy loss (Triplet Entropy Loss) and tests it on South-African NCHLT Speech corpus [2] to achieve 81% accuracy with a rather large Imagenet-pretrained Densenet-121 model [20].

In a more traditional linguistic setting, Sarthak et al. [34] explore 1D-ConvNet that auto-extracts and classifies features from raw audio input and 2D-ConvNet architectures, and enhance the performance of these approaches by utilizing Mixup augmentation of inputs and attention mechanism. They achieve 93.7% and 95.4% overall accuracy on a six language (En, Fr, De, Es, Ru, It) dataset with overlapping phonemes based on the VoxForge [38] dataset.

Valk and Alumae [36] have collected a 107-language dataset from YouTube and then used language embedding models followng the x-vector paradigm [35]. During training on-the fly data augmentation using AugMix [11] was applied by randomly distorting the training data using a mix of reverberation and noise augmentation. For frame-level feature extraction, Resnet34 [10] architecture was used with the basic convolutional blocks with residual connections replaced by squeeze-and-attention modules [12][41], and for the temporal pooling, multi-head attention similar to the one described in [19]. Using the system described above, Valk and Alumae achieved 7.1% average error rate on VoxLingua107 dataset.

## 1.2 Our contribution

Our contribution in this memo is threefold:
1. We propose a very practical LID model based on 1D time-channel separable convolutions. This type of models is very fast [15], trains well and is widely used in the industry.
2. We set a SOTA for the dataset [20] using this model.
3. We state and substantiate the hypothesis that whenever the dataset is diverse enough so that the other classification factors, like gender, age etc. are well-averaged, the confusion matrix for LID system bears the language similarity measure.

## 2 Model architecture

Similar to [14], the model is based on 1D time-channel separable convolutions, namely, the QuartzNet ASR architecture [15] comprising of an encoder and decoder structures. We have used cross-entropy loss function in this work.

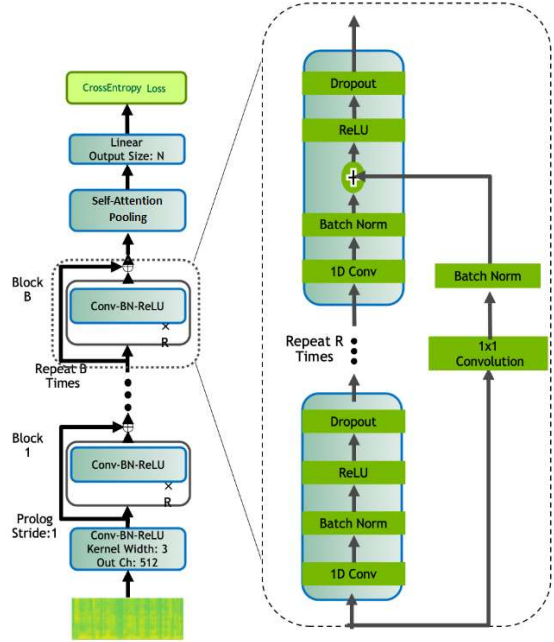

Figure 1. The model architecture

### 2.1 Encoder

The encoder used is shown in Figure 1 and is a QuartzNet [15] BxR model with B blocks, each with R sub-blocks. The first block is fed with MFSC coefficients vector of length 40. Each sub-block applies the following operations [15]:
- a 1D convolution,
- batch norm,
- ReLU, and
- dropout.

All sub-blocks in a block have the same number of output channels. These blocks are connected with residual connections [15]. We use QuartzNet 15*5, with 512 channels. All the convolutional layers have stride 1 and dilation 1 [14]. No hyperparameter optimization was performed.

## 2.2 Self-attentive pooling decoder

Similar to [4][5][6], we agree that not all frames contribute equally to the utterance level representation, Thus we use a self-attentive pooling (SAP) layer introduced in [5][41] to pay more attention to the frames that are more important.

Namely, we first feed the frame level feature maps $\{x_1, x_2, \cdots, x_L\}$ into a fully-connected layer to get a hidden representation

$$h_t = tanh(Wx_t + b)$$

Then we measure the importance of each frame as the similarity of $h_t$ with a learnable context vector μ and get a normalized importance weight $w_t$ through a softmax function [5].

After that, the utterance level representation $e$ can be generated as a weighted sum of the frame level feature maps based on the learned weights:

$$e = \sum_{t=1}^{T} w_t x_t$$

## 3 Experiments

### 3.1 Datasets and tasks

Lowresource-lang-eval dataset [20] contains 19k utterances from various speakers, each utterance up to 20 second long. Each of these utterances in the training dataset belongs to one of 23 classes corresponding to the languages of Russia, mostly Siberian, like "krl", "evn", "koi-yzv", "mrj", etc.

Train and validation subsets were generated by splitting the dataset randomly in 80%-20% proportion. The test dataset additionally contains languages that were not present in the training dataset. Test dataset consists of 10445 samples, 4702 of them spoken in languages that are present in the training data.

In our experiments, we have considered the following tasks for Lowresource-lang-eval dataset:
- Recognition of 23 languages
- Recognition of 4 language groups
- Recognition of 2 language families.

For all tasks, we have measured top-1 classification accuracy. Due to the submission deadline constraints, the requested function to determine languages that were not present in the training dataset was not implemented.

VoxForge is an open-source speech corpus that primarily consists of samples recorded and submitted by users using their own microphone. This results in significant variation of speech quality between samples. Similar to [34], we selected 1,500 samples for each of six languages. Out of 1,500 samples for each language, 1,200 were randomly selected as training dataset for that language and rest 300 as validation dataset using k-fold cross-validation. Thus, we have trained our model on 7,200 samples and validated it on 1800 samples comprising six languages.

### 3.2 Optimization and training process

We have used the attention vector size of 256. Models were trained until they reached a plateau on a validation set. Training was performed using the Stochastic Gradient Descent optimizer with the initial learning rate of 0.005 and cosine annealing decay to 1e-4. SpecAugment [27] was used for augmentation.

## 4 Results and Discussion

The system described above allowed us to achieve the following results on Lowresource-lang-eval (see Table 1), that happened to be a winning entry for the competition in the terms of language and language group identification accuracy:

| Task | Top-1 accuracy, % (test) | Top-1 accuracy, % (validation) |
|---|---|---|
| Language identification | 6.29 | 80.31 |
| Language group identification | 34.71 | 85.87 |
| Language family identification | 61.71 | 89.51 |

Table 1: Test and validation accuracy

The low test accuracy, especially, related to the results on the validation set, is most likely explained by the fact that the test set contains extra languages not present in the training and validation sets. Determining the unknown languages can be achieved using a sort of metric embedding and kNN / voting decoder a-la, for example, [39], where a similar structure (triplet loss based embedding + kNN decoder) was used for keyword spotting, which is, basically, another utterance classification task. This is in our plans for further research. On the other hand, it is interesting that the language family and even the group was predicted with a rather high accuracy.

It is pretty instructive to analyze the language confusion matrices. We have split the matrix into two pieces: one for the languages that were present in the training set (Figure 2), and the other – for the ones that were not in it (Figure 3).

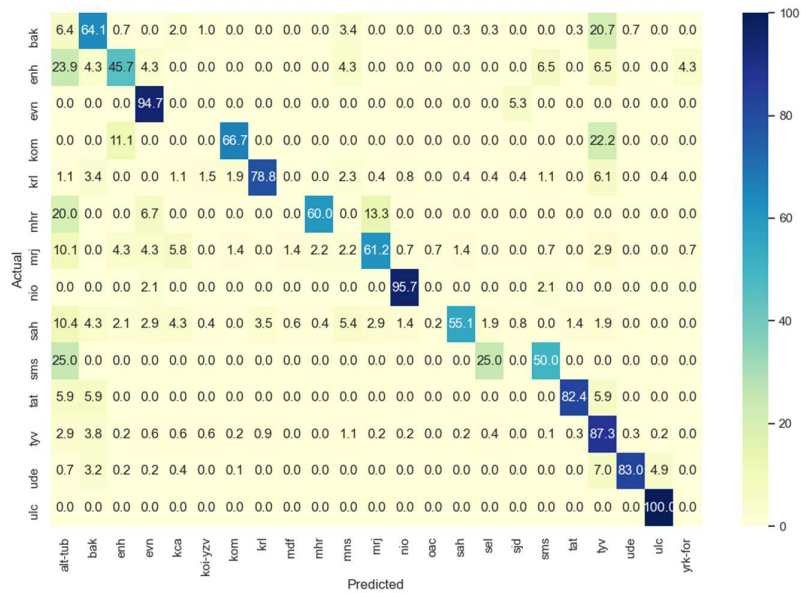

Figure 2: Confusion matrix for the languages present in the Lowresource-lang-eval training set

Languages present in the training set mostly determine well. Tuvan and Tuba happened to be the languages most commonly misattributed by our model, so this is a model deficiency that does not bear any valuable information. On the other hand, there are too few examples of Komi, Meadow Mari and Saami in the test dataset, so any confusions of these languages are statistically insignificant. Mansi is an interesting example of language to which only languages absent in the training set are misattributed. This means that it separated well from the known, but not from unknown to the model languages.

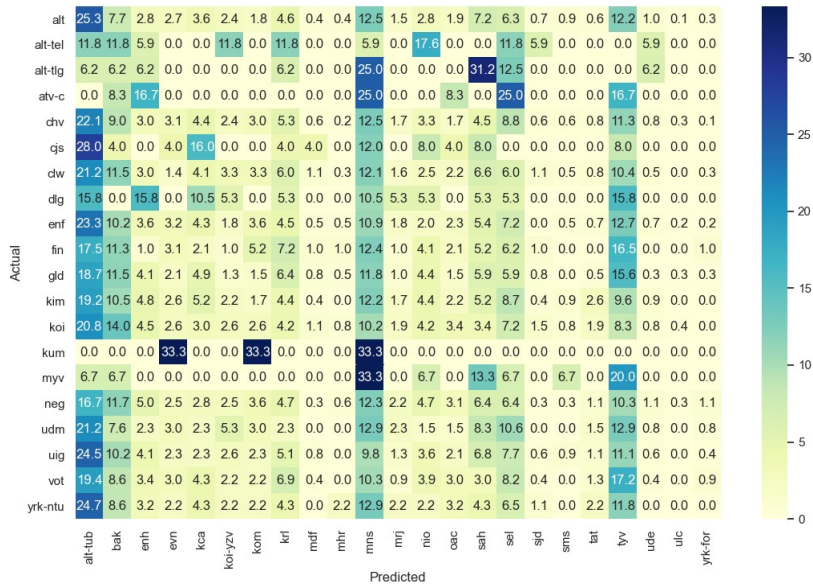

Figure 3: Confusion matrix for languages not present in the Lowresource-lang-eval training set

On the VoxForge dataset [38], we have achieved 92.5% accuracy. This is somewhat lower than the results reported in [34], but still within the significance bounds, given the varied speech domains and small sample selection from the bigger corpus. The confusion matrix is depicted below in Figure 4.

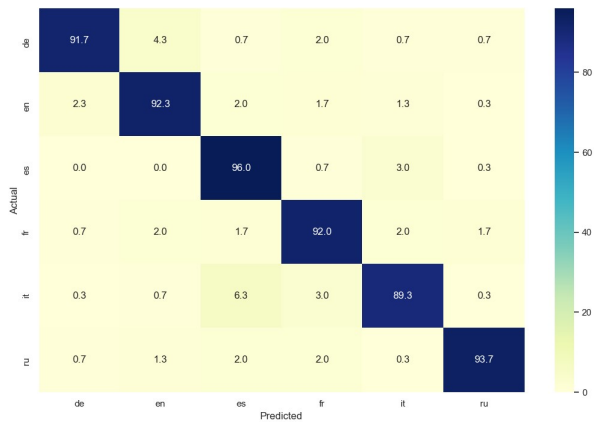

Figure 4: Confusion matrix for VoxForge dataset

Quite unlike Lowresource-lang-eval, in this dataset the greatest confusion is between the closely related languages: 6.3% between Spanish and Italian that belong to Romance group, and 4.3% between English and German that belong to Germanic group. On the other hand, more distant languages like Italian, on one side, and German or Russian on the other, are hardly confused. From yet another viewpoint, all these languages belong to the Indo-European family, so are not that distant, and had historically strong ties, like Russian and French (from which many of contemporary Russian words have been adopted in 18$^{th}$-19$^{th}$ centuries). The latter can explain the fact that the language most confused with Russian is French.

Thus, we can hypothesize that confusion between languages in a more diverse dataset can bear more linguistic information about the language relations – this is also somehow unsurprising, because in a more diverse dataset other classification factors, like gender, age etc. are better averaged. This is yet another topic for potential further research.

## Acknowledgments

The authors are grateful to:
- colleagues at NTR Labs Machine Learning Research group for the discussions and support;
- anonymous reviewers who suggested important improvements to this memo.